# Measuring laser carrier-envelope phase effects in the noble gases with an atomic hydrogen calibration standard

Champak Khurmi[1,2], W. C. Wallace[1,2], Satya Sainadh U[1,2], I. A. Ivanov[3,4], A. S. Kheifets[3], X. M. Tong[5,6], I. V. Litvinyuk[1,2], R. T. Sang[1,2], and D. Kielpinski[1,2]

[1]Australian Attosecond Science Facility, Griffith University, Nathan, Qld, Australia
[2]Centre for Quantum Dynamics, Griffith University, Nathan, Qld, Australia
[3]Research School of Physics and Engineering, Australian National University, Canberra ACT 0200, Australia
[4]Center for Relativistic Laser Science, Institute for Basic Science, Gwangju 500-712, Republic of Korea
[5]Center for Computational Sciences, University of Tsukuba, Ibaraki, Japan
[6]Faculty of Pure and Applied Sciences, University of Tsukuba, Ibaraki, Japan

We present accurate measurements of carrier-envelope phase effects on ionisation of the noble gases with few-cycle laser pulses. The experimental apparatus is calibrated by using atomic hydrogen data to remove any systematic offsets and thereby obtain accurate CEP data on other generally used noble gases such as Ar, Kr and Xe. Experimental results for H are well supported by exact TDSE theoretical simulations however significant differences are observed in case of noble gases.

The last decade has seen rapid development in few-cycle laser technology and carrier envelope phase (CEP) stabilised laser systems are now routinely used in many research labs. The CEP of a laser pulse is generally defined as the phase of the carrier frequency with respect to the intensity envelope. An important aspect of the few-cycle laser pulses ($\lambda_{central}$ = 780 nm, 1 optical cycle ~ 2.6 fs) is that the CEP determines the time evolution of the electric field of the laser pulse and also affects the processes initiated by the laser pulse when interacting with the matter. Therefore the ability to precisely measure the CEP of few-cycle laser pulses is very important for diverse scientific applications such as high harmonics generation [1,2], above-threshold ionisation (ATI) [3], attosecond pulse generation [4,5], coherent control of molecular dynamics [6-8] and attosecond ionisation under the influence of strong laser fields [9]. Experiments by G. Paulus [10] have shown that the CEP of few-cycle laser pulses can be tagged by using Xe atoms as target species. But questions remain about systematic CEP phase offsets in such measurements, since the accuracy of the available theoretical models is not well characterized [11]. However, in case of atomic hydrogen exposed to an intense few-cycle laser pulse, the Time Dependent Schrodinger Equation (TDSE) can be solved numerically with high precision and provides very reliable calibration standard [12-14]. Here, we present the experimental evidence of the measurement of CEP of few-cycle laser pulses in the noble gases by using atomic hydrogen as the calibration standard. In the case of H, we find that the experimental results are well supported by ab-initio theoretical simulations however for multi-electron species such as Ar, Kr and Xe, in the Single Active Electron (SAE) approximation [15] differ significantly from experimental results.

Figure 1 shows a schematic of the experiment, a commercial *Femtosecond Compact Pro* laser system with CEP stability is used to generate ~6 fs laser pulses with 780 nm central wavelength. An additional f-2f interferometer (from Menlo Systems) is used to establish CEP feedback and locking near the experimental end station to control slow CEP drifts. A set of matched fused silica wedges (from MolTech GmbH Berlin, 1 mm lateral translation = 1.25 radians phase shift) is used to vary the CEP of the laser pulses.

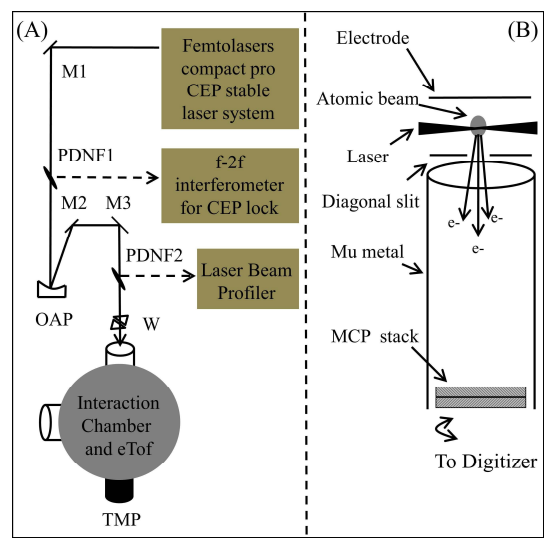

Figure 1. (A) Experimental setup for CEP resolved experiments. M1-M3: Reflective mirrors, PNDF: Pellicle Neutral Density Filters, OAP: Off-Axis Parabolic Mirror, W: Fused Silica Wedge, TMP: Turbo Molecular Pump (B) Electron Time of Flight Spectrometer.

The laser beam is focused into the interaction chamber by using an Off-Axis Parabolic (OAP, focal length = 750 mm) mirror and interacts with the atomic H beam in the interaction chamber. Electrons generated from this interaction are detected by an Electron Time of Flight detection system (EToF). This EToF spectrometer (fig. 1 (B)) is enclosed in μ-metal to provide shielding from stray magnetic fields. Electrons with different kinetic energies are generated from laser and atom interaction and travel in a field free region to a micro-channel plate (MCP). Each electron gives rise to a temporally resolved voltage peak which is recorded using an Analog to Digital conversion card (from Agilent, model number U1084A). Fig. 2 shows the CEP averaged electron energy spectra of different atomic species (H, Ar, Kr and Xe) at two laser peak intensities, namely at $1.2\times10^{14}$ W/cm$^2$ and $2.5\times10^{14}$ W/cm$^2$. The electron energy spectra for each atomic species have been offset for the sake of clarity. For CEP resolved experiments, a motorized fused silica wedge is used to vary the laser CEP over a range exceeding $2\pi$ radians.

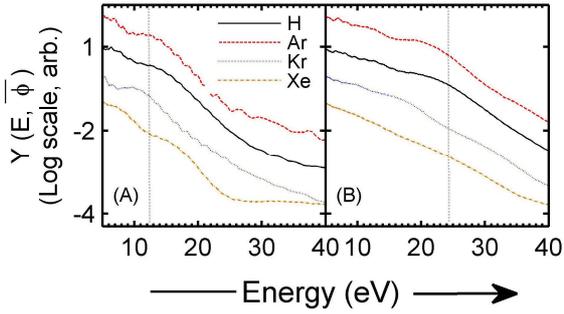

Figure 2. CEP averaged electron energy spectra of different atomic species namely H (solid line, black), Ar (dashed line, Red), Kr (dash-dotted line, Blue) and Xenon (dotted line, Orange) at two different intensities. Electron energy spectra for different atomic species are offset for the sake of clarity. (A) $1.2\times10^{14}$ W/cm$^2$ and (B) $2.5\times10^{14}$ W/cm$^2$. Dotted grey lines represent 2U$_p$ point.

The electron energy spectrum is collected at each wedge position (integration time 90 sec) representing a particular laser CEP point. The CEP-dependent electron spectrum is denoted $Y(E,\phi)$, where E is the electron kinetic energy. As seen from Fig. 2, $Y(E,\overline{\phi})$ varies over a wide range, so we parametrize the laser CEP effects by the normalized quantity

$$S(E,\phi) = \frac{Y(E,\phi) - Y(E,\overline{\phi})}{Y(E,\overline{\phi})} \quad (1)$$

$S(E,\phi)$ measures the CEP effect at E relative to the average electron yield at E.

We obtain theoretical simulations for H from numerical integration of the three-dimensional TDSE. These simulations are therefore extremely reliable. For multielectron systems such as Ar, Kr and Xe, the theoretical simulations are based on SAE approximation. The ATI spectra of the rare gas atoms are calculated by solving the time-dependent Schrodinger equation (TDSE) with the generalized spectrum in the energy representation [15,16] under the single-active electron approximation. We use the model potentials [17] obtained by density functional theory with the self-interaction correction [18], which gives the atomic ionization potentials. The numerical convergence has been checked by varying the number of partials, grid points and so on. Focal volume integration is performed on all simulations for comparison with experimental data.

Figures 3 and 4 show the experimental data and theoretical simulations for H, Ar, Kr, and Xe. For ease of viewing, both data and theory are smoothed with respect to energy using a Gaussian filter with full-width at half-maximum (FWHM) of 1.5 eV and the results over the CEP range $0 < \phi < 2\pi$ are replicated over the range $2\pi < \phi < 4\pi$. The Gaussian filter with 1.5 eV FWHM was chosen based on the ETof detector resolution. The close match between experiment and theory for H demonstrates that our experimental results are reliable. The data for the noble gases are taken under identical conditions in the same apparatus, so they are expected to be similarly reliable. We can therefore assign an absolute CEP to the data on the noble gases free of systematic errors. Our calibrated data on the noble gases Ar, Kr, and Xe show that the SAE-based predictions are only qualitatively accurate.

For quantitative analysis of the CEP effects, we bin the data and simulations with respect to energy. A bin width of 5 eV was found to represent a good compromise between signal-to-noise ratio and energy resolution. The data in the bin centred on energy $E$ is denoted $B_E(\phi)$, and we fit such data to a sinusoidal function for the CEP effects

$$B_E(\phi) = \sin(\phi + \phi_0) \quad (2)$$

with the "offset" phase $\phi_0$ as a fit parameter. This offset phase obtained from experimental data are plotted for each atomic species in fig. 5 (A) and (B) for $1.2\times10^{14}$ W/cm$^2$ and $2.5\times10^{14}$ W/cm$^2$ respectively. The difference between the estimated experimental and theoretical offset phase is shown in fig. 5 (C) and (D) for both laser intensities respectively. As expected, the experimental offset phase for H is very close to the theoretical prediction. For the noble gases, the quantitative analysis shows two principal points of difference between experiment and theory. First, in the energy region above $2U_p$ (where $U_p$ is the ponderomotive energy), the observed CEP effects depend on energy, contrary to the SAE simulations. Second, the SAE simulations display a large systematic offset in CEP relative to the actual values obtained from the TDSE. Notably, only the use of the H calibration can reveal this offset, since SAE (for multielectron atoms) and direct TDSE (for H) can never be directly compared by purely theoretical means. Furthermore at $2.5\times10^{14}$ W/cm$^2$ laser peak intensity, the Keldysh parameter is <1 for all atomic species under investigation. This represents the onset of the photoionization tunnelling regime where the potential barrier of the atom is lowered by the intense laser electric field and the electron can tunnel through the Coulomb potential barrier. When the direction of the laser electric field reverses after the half cycle, the ionised electron may recombine with the parent ion to give rise to high harmonics generation (HHG) or collide with the parent ion to cause non-sequential double ionisation. As such the true interpretation of HHG spectra or the kinetic energy of emitted electrons at higher intensities will depend on the accurate measurement of the CEP of few-cycle laser pulses.

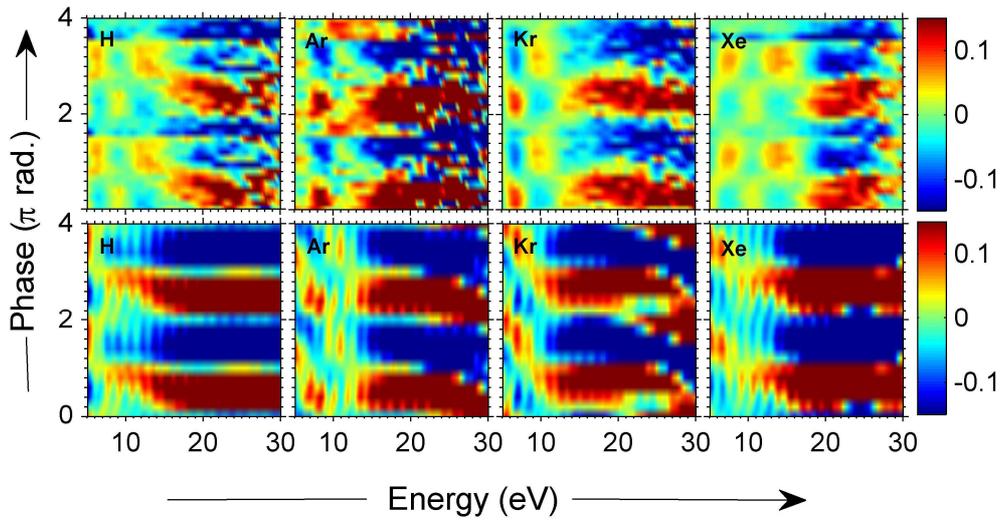

Figure 3. CEP maps for different atomic species at $1.2\times10^{14}$ W/cm$^2$. Top panel (left to right): experimental results for H, Ar, Kr and Xe. Bottom panel (left to right) theory data. In case of H, exact TDSE simulations are used whereas for Ar, Kr and Xe, theoretical simulations are based on SAE approximations.

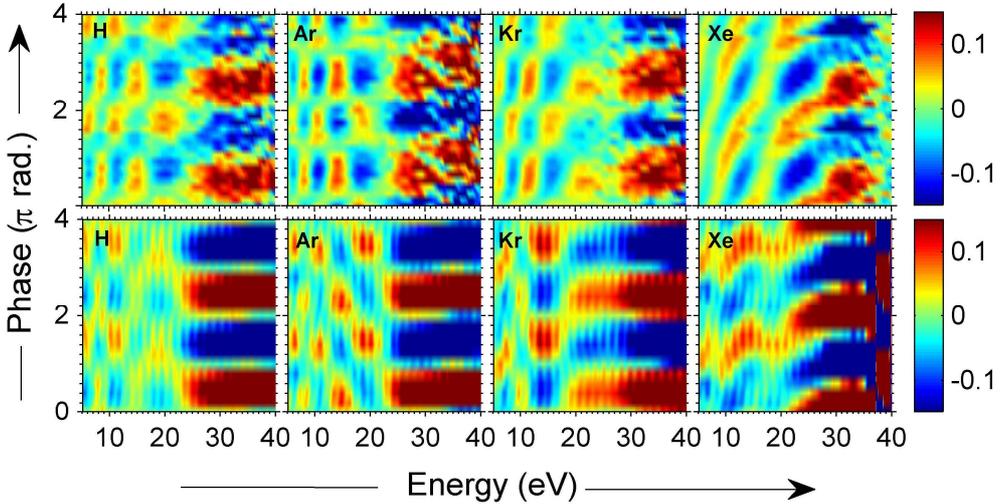

Figure 4. CEP maps for different atomic species at $2.5\times10^{14}$ W/cm$^2$. (Top panel, left to right) CEP resolved experimental results for H, Ar, Kr and Xe. (Bottom panel, left to right) Theoretical simulations, in case of H, exact TDSE simulations are used whereas for Ar, Kr and Xe, theoretical simulations are based on SAE approximations.

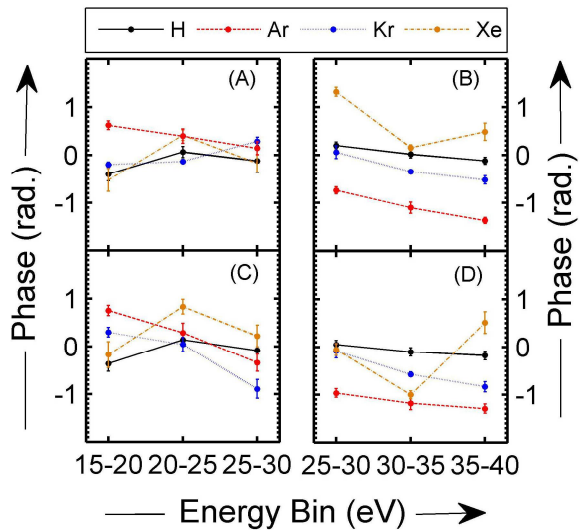

Figure 5. Experimental phase offset (bin width = 5 eV) for different atomic species at (A) $1.2\times10^{14}$ W/cm$^2$ and (B) $2.5\times10^{14}$ W/cm$^2$. Difference between phase offset from experiment and theoretical simulations for (C) $1.2\times10^{14}$ W/cm$^2$ and (D) $2.5\times10^{14}$ W/cm$^2$. Lines are guide to the eye.

In conclusion, our experimental results clearly depict that one cannot rely only on approximate theoretical methods such as SAE to accurately calibrate the CEP of the few-cycle laser pulses using noble gases. H CEP data provides an extremely reliable calibration standard and removes any systematic errors due to the experimental setup and provides a reliable calibration for noble gas species. It is also evident that the interaction of laser pulses with multielectron systems shows significant dependence on CEP phase at higher intensity where the Keldysh parameter is <1. Recent experiments [19] on He atoms using few-cycle laser pulses also suggest that at higher intensities (2-4×10$^{14}$ W/cm$^2$) the electron correlation effects of bound state electrons play significant role in determining the time resolved absorption spectra of auto-ionizing states. This work lays the foundations for further experiments on H and noble gases in this regard.

Acknowledgements:

This work was supported by the United States Air Force Office of Scientific Research under Grant FA2386-12-1-4025. D.K. was supported by ARC Future Fellowship FT110100513. I. A. I. was supported by the ARC Discovery Grant DP 120101085.